\documentclass[12pt]{article}
\usepackage{amsmath}

\allowdisplaybreaks
\begin{document}
\thispagestyle{empty}
\vspace*{-15mm}
%----------
\baselineskip 10pt
\begin{flushright}
\begin{tabular}{l}
{\bf EPHOU-03-002}\\
%{\bf OCHA-PP-yyy}\\
%{\bf May 2003}\\
{\bf hep-th/0305106}
\end{tabular}
\end{flushright}
\baselineskip 24pt
\vglue 10mm
%%%%%%%%%%%%%%%% Title %%%%%%%%%%%%%%%%%%% 
\begin{center}
{\LARGE\bf Supersymmetric Matrix Model}\\
{\LARGE\bf on Z-Orbifold}
\vspace{7mm}

\baselineskip 18pt
{\bf Akiko MIYAKE}\footnote{miyake@particle.sci.hokudai.ac.jp} 
%{\bf and Akio SUGAMOTO}\footnote{sugamoto@phys.ocha.ac.jp} 
\vspace{2mm} 

{\it Department of Physics, Faculty of Science}\\
{\it Hokkaido University}\\
{\it Sapporo, 060-0810, Japan}\\
%\vspace{2mm}
%${}^2${\it Department of Physics, Ochanomizu University}, \\
%{\it 2-1-1, Otsuka, Bunkyo-ku, Tokyo 112-8610, Japan }\\
\vspace{10mm}
\end{center}
%%%%%%%%%%%%%%%% Abstract %%%%%%%%%%%%%%%
\begin{center}
{\bf Abstract}\\[7mm]
\begin{minipage}{14cm}
\baselineskip 16pt
\noindent
%%%%%---------------------------------
We find that the IIA Matrix models defined on the non-compact $C^3/Z_6$,
 $C^2/Z_2$ and $C^2/Z_4$ orbifolds preserve supersymmetry where the
 fermions are on-mass-shell Majorana-Weyl fermions. In these examples
 supersymmetry is preserved both in the orbifolded space and in the
 non-orbifolded space at the same time. The Matrix model on $C^3/Z_6$
 orbifold has the same ${\cal N}=2$ supersymmetry as the case of
 $C^3/Z_3$ orbifold which was pointed out previously.
 On the other hand the Matrix models on $C^2/Z_2$ and $C^2/Z_4$
 orbifold have a half of the ${\cal N}=2$ supersymmetry. We further find 
 that the Matrix model on $C^2/Z_2$ orbifold with a parity-like
 identification preserves ${\cal N}=2$ supersymmetry.

%%%%%----------------------------------
\end{minipage}
\end{center}

%%%----------------------------------
%%%
%%%
%%% Main body of the paper
%%%
%%%
%%%----------------------------------
\newpage
\baselineskip 18pt
\def\thefootnote{\fnsymbol{footnote}}
\setcounter{footnote}{0}

\section*{Introduction}

The dynamics of the extended objects, D-branes \cite{Polchinski}, the key 
ingredients to understand the nonperturbative effect of strings and their 
property of duality, can be described in terms of the
dimensionally reduced Supersymmetric Yang-Mills
(SYM) Theory \cite{Witten}.  In particular, the dynamics of the
simplest point-like D-particles may play the fundamental role \cite{BFSS} 
and it is described by the matrix description of the 11d super-membrane 
theory \cite{Hoppe}.  On the other hand the dynamics
of D-particles can be viewed as the SYM theory dimensionally reduced to one 
temporal dimension, that is the quantum mechanics of the SUSY Matrix
model.  Here, the  bosonic fields of the SUSY Matrix
model describe  the noncommutative bosonic coordinates 
consisting of
$N$ D-particles. They are represented by $N\times N$ hermitian matrices. 
We call this the 11d M-theory.

The 11d M-theory compactified on a simple orbifold space, $S^1/Z_2$,
was shown to be connected with the 10d $E_8\times E_8$ heterotic string
theory \cite{Horava-Witten}. While the 6d compactification on the
Calabi-Yau manifold of the usual 10d heterotic string theory was given by 
\cite{Candelas}, and the compactification on the various orbifold spaces 
was studied in \cite{Nilles},\cite{Katsuki-Kobayashi},\cite{Senda-Sugamoto}.

The type IIA Matrix models compactified on the $Z_2$ orbifold spaces were
explicitly investigated by
\cite{Danielsson-Ferreti}, \cite{Kachru-Silverstein}, \cite{S.J.Rey},
\cite{Banks-Motl}, \cite{Kachra-Lawrence-Silverstein}, and others.
On the other hand, the type IIB Matrix model was formulated by \cite{IKKT}, 
and the type IIB $USp(2k)$ Matrix model compactified on the $Z_2$
orbifold spaces was investigated by \cite{Itoyama}.

In order to understand the realistic world in terms of the string dynamics 
or the M-theory, we need to find Matrix models defined on the flat 4d 
space-time and non-flat 6d spaces.

Explicit examples of the Matrix models defined on the 4d flat space-time
and non-compact 6d orbifold spaces are
demonstrated by \cite{Iso}, \cite{Miyake-Sugamoto}.

In the supersymmetric Matrix model defined on $C^3/Z_3$
orbifold \cite{Miyake-Sugamoto}, the number of supersymmetry is reduced to 
${\cal N}=2$ .
In this paper we generalize the previous study on the Matrix model on 
$C^3/Z_3$ to more general orbifold spaces.  We first formulate the 
Matrix model on the  generic $C^i/Z_n$ orbifold, by using the 't Hooft 
algebra for $SU(n)$ monopoles.
After examining carefully the structure of fermions satisfying the 
Majorana conditions, the  Supersymmetric IIA Matrix models on the 
non-compact $C^i/Z_n$ orbifold spaces are found to be limited in number. 
Examples of these supersymmetric Matrix models are given by 
$C^3/Z_6$, $C^2/Z_2$ and 
$C^2/Z_4$ orbifold.   In these examples supersymmetry is preserved both in 
the orbifolded space and in the (non-orbifolded) flat space at the same 
time. The number of supersymmetry, however,  depends on the models.  The 
Matrix model on $C^3/Z_6$
  orbifold has the same ${\cal N}=2$ supersymmetry as the case of
  $C^3/Z_3$ orbifold investigated previously. On the other hand the
  Matrix models 
on $C^2/Z_2$ and $C^2/Z_4$ orbifolds have the same number as ${\cal
N}=1$ supercharges.

We further study the case in which the parity symmetry (the reverse of the 
space coordinates) is supplemented to the $Z_2$ symmetry, where the
parity-like $Z_2$ 
identification is defined. We then find that the Matrix 
model on $C^2/Z_2$ orbifold with this parity-like identification differs 
from the usual $C^2/Z_2$ orbifold, and the former model  preserves ${\cal 
N}=1$ supersymmetry both in the orbifolded space and non-orbifolded space, 
and has ${\cal N}=2$ supersymmetry parameters in the total space.

\section*{IIA Matrix model}

We treat $nN$ D-particles. The action of the IIA Matrix model is given by
\begin{eqnarray}
S=\frac{1}{2g^2}\int dt~{\rm
 Tr}\left\{(D_t X^I)^2 +\frac{1}{2}
[X^I, X^J]^2 -ig^2 \Theta^T D_t \Theta -g^2\Theta^T {\tilde \Gamma}^I 
[X_I, \Theta]\right\},
\end{eqnarray}
where $I,~J =1,\cdots ,9$ label the Minkowski space-time indices.
$\Gamma^I$ is the 10 dimensional gamma matrices with
${\tilde \Gamma}^I=\Gamma^0\Gamma^I$,
$g$ is the Yang-Mills coupling and 
$D_t$ is the covariant derivative defined by
\begin{eqnarray}
D_t =\partial_t -i [X_t,~~].
\end{eqnarray}
The bosonic fields $X_I$ and the fermionic fields $\Theta$ are 
$nN\times nN$ hermitian matrices. The fermionic fields $\Theta$ are
given by the Majorana-Weyl fermions and have sixteen degrees of 
freedom on the mass shell. These fields depend on temporal dimension.
The diagonal parts of the bosonic fields $X_{I}$ are given by the
solutions of equations of motion and represents the classical
coordinates.
The other parts of the bosonic, 
fermionic and ghost fields have the fluctuating fields which represent
interaction among 
the $nN$-body D-particles. 

The supersymmetric transformation of IIA Matrix model is described 
as follows: 
\begin{eqnarray}
%\left\{\begin{array}{lll}
\delta X_t &=&ig\epsilon^T \Theta,\\
\delta X_I &=&ig\epsilon^T {\tilde \Gamma}_I \Theta,\\
\delta \Theta &=&\frac{1}{g}\left({\tilde \Gamma}^I (D_t X_I)
\epsilon -\frac{i}{2}\Gamma^{IJ}[X_I,X_J]\epsilon\right) +\xi,
%\end{array}\right.
\end{eqnarray}
where $\Gamma^{IJ}=\frac{1}{2}[ \Gamma^{I}, \Gamma^{J} ]$.
$\epsilon$ is the supersymmetric transformation parameter of 
${\cal N}=1$ SYM theory and
$\xi$ is the translation parameter of the fermions.
This model has ${\cal N}=2$ SUSY.

\section*{IIA Matrix model on the $C^i/Z_n$ Orbifold}

It is convenient to use the complex notations
for the orbifolded dimensions.
The complex coordinates $Z_j$ is defined as follows:
\begin{eqnarray}
Z_i\equiv X^{2i}+iX^{2i+1},
\end{eqnarray}
where $i=2,3,4$ for three 
complex spaces $C^3$, $i=3,4$ for the two 
complex spaces $C^2$ and $i=4$ for the one complex space $C^1$. 
We keep the flat space-time coordinates at least 
as 4 dimensions, and take the orbifold spaces for the other 
dimensions. 
Let us impose a $Z_{n}$ symmetry about the complex coordinates $Z_{j}$
and obtain the $Z_{n}$-orbifold.
The complex coordinates $Z_{j}$ have $Z_n$ symmetry under
\begin{eqnarray}
Z_j\simeq \omega Z_{j},
\end{eqnarray}
where $\omega =e^{2\pi i/n}$.

To impose the $Z_n$ symmetry on the $nN\times nN$ matrices, 
we use the 't Hooft matrices; $U$ and $V$:
\begin{eqnarray}
U=\left(\begin{array}{ccccc}
1 & & & & \\
 & \omega & & & \\
 & & \ddots & & \\
 & & & \ddots & \\
 & & & & \omega^{n-1}
\end{array}\right)_{nN\times nN},\quad
V=\left(\begin{array}{ccccc}
0 & & & & 1 \\
1 & 0 & & & \\
 & 1 & \ddots & & \\
 & & \ddots & \ddots & \\
 & & & 1 & 0 
\end{array}\right)_{nN\times nN},
\end{eqnarray}
where the block matrices is $N\times N$ matrices. $U$ and $V$ satisfy 
$UU^{\dag}=U^{\dag}U=U^n=1$,
$VV^{\dag}=V^{\dag}V=VV^{T}=V^{T}V=V^n=1$ and the following relation:
\begin{eqnarray}
UV=\omega VU.
\end{eqnarray}

We start with the $nN$-body system of the D-particles in the complex 
space(s) $Z_{j}$. And we divide
the complex space(s) $Z_{j}$ into $n$ equal parts. Under the $Z_n$
symmetry, $N$ 
D-particles are distributed into a small part of the same size.
There are $n$ mirror images in the complex space(s) $Z_j$.
Let us use a $SO(nN)$ group $V$ and impose the $Z_n$ invariance 
on the bosonic and fermionic fields 
\begin{eqnarray}
X_{/\!\!/}^{\mu}&=& VX_{/\!\!/}^{\mu}V^{\dag},\label{flat spaces}\\
Z_{j}&=& \omega_j VZ_j V^{\dag},\label{orbifold spaces}\\
\Theta &=& {\hat \omega}V\Theta V^{\dag}, \label{orbifold ferumions}
\end{eqnarray}
where 
\begin{eqnarray}
\omega_j&=&\exp\left(2\pi i\frac{n_j}{n}\right),\nonumber\\
{\hat \omega}&=& \exp \left(2\pi i\sum_{j=2}^{4}\frac{n_j}{n}
b_{j}^{\dag}b_{j}\right).
\end{eqnarray}
$n_{j}$ is an integer. ${\hat \omega}$ represents the $Z_n$ symmetry for
10 dimensional Majorana-Weyl spinors on the mass shell. The 16 
Majorana-Weyl spinors can be represented by using the raising and 
the lowering operator; $b_{\mu}$ and $b_{\mu}^{\dag}$.
$b_{\mu}$ are defined as follows: 
\begin{eqnarray} 
%\left\{\begin{array}{l}
b_{0}&=&\frac 12 (\Gamma^{1}-\Gamma^{0}),\\ 
b_{j}&=&\frac 12 (\Gamma^{2j}
-i\Gamma^{2j+1})~,\quad j=1,\ldots ,4. 
%\end{array}\right
\end{eqnarray}
The Gamma matrices $\Gamma^{\mu}$ satisfy the Clifford algebra 
$\{ \Gamma^{\mu},\Gamma^{\nu} \}=2\eta^{\mu\nu}$, where 
$( \Gamma^{0})^{2}=-1$ and $(\Gamma^{j})^{2}=1$. 

The 16 Majorana-Weyl spinors on the mass shell are written as
\begin{eqnarray}
\Theta^a =\left(\begin{array}{c}
\psi^{a}|0\rangle\\
\psi^{a*}|0\rangle\\
\psi^{[ij]a}b_i^{\dag}b_{j}^{\dag}|0\rangle\\
\psi^{[ij]a*}b_i^{\dag}b_{j}^{\dag}|0\rangle
\end{array}\right),
\end{eqnarray}
where $i=1,\cdots ,4$. The superscript $a$ denotes the number of the 
supersymmetry; $a=1,2$. The ground state of the spinors is defined
as follows:
\begin{eqnarray}
|0\rangle\equiv |-,-,-,-,-\rangle ,
\end{eqnarray}
where the first minus sign is fixed on the mass shell and  the others 
can be changed into the plus sign using the raising operator $b_i^{\dag}$.

We impose the Majonara condition:
\begin{eqnarray}
B\Theta = \Theta^*,
\end{eqnarray}
where $B=\Gamma^3\Gamma^5\Gamma^7\Gamma^9$.
From this condition, we obtain
\begin{eqnarray}
\sum_{i=2}^{4}\frac{n_j}{n}=integer ~\mbox{or} ~half\!\!-\!\!integer.
\label{Majorana condition}
\end{eqnarray}

\section*{IIA Matrix model on the special $C^i/Z_n$ Orbifold}

In this section, we treat the special $C^i/Z_n$ orbifolded spaces where
$n_j$ does not depend on the subscript $j$. In other words, the phases 
$\omega_j$ are equal in the orbifolded complex spaces;
${n_2}={n_3}={n_4}$ for the $C^3/Z_n$ orbifold,
${n_3}={n_4}$ for the $C^2/Z_n$ orbifold and
${n_4}$ for the $C^1/Z_n$ orbifold.
From  eq. (\ref{Majorana condition}), The combination 
$(\frac{n_2}{n},\frac{n_3}{n},\frac{n_4}{n})$ is given as follows:
\begin{eqnarray}
\left(\frac{1}{2},\frac{1}{2},\frac{1}{2}\right),\left(\frac{1}{3},
\frac{1}{3},\frac{1}{3}\right),\left(\frac{1}{6},\frac{1}{6},
\frac{1}{6}\right),
\quad \mbox{on}~C^3/Z_{2,3,6},\nonumber\\
\left(1,\frac{1}{2},\frac{1}{2}\right),
 \left(1,\frac{1}{4},\frac{1}{4}\right),\quad
\mbox{on}~C^2/Z_{2,4},\nonumber\\
\left(1,1,\frac{1}{2}\right),\quad \mbox{on} ~C^1/Z_2.
\label{special C^3/Z_{2,3,6}, C^2/Z_{2,4}, C^1/Z_2}
\end{eqnarray}
Using eqs. (\ref{flat spaces}), (\ref{orbifold spaces}) and 
(\ref{orbifold ferumions}), we can get the bosonic fields and
the fermionic fields on the special $C^i/Z_n$ orbifolds.
We decompose the $nN\times nN$ matrices into the sum of the tensor
products $(N\times N ~matrices)\otimes (n\times n ~matrices)$ and then
we obtain
\begin{subequations}
\begin{eqnarray}
X_{/\!\!/}^{\mu}
&=&H_1^{\mu}\otimes 1+A_1^{\mu}\otimes V^{\dag}+ A_{2}^{\mu}\otimes 
(V^{\dag})^{2}+\cdots +A_{m}^{\mu}\otimes (V^{\dag})^{m}\nonumber\\
&&+A_{m}^{\dag\mu}\otimes V^{m}+\cdots + A_{2}^{\dag\mu}\otimes V^2 +
A_{1}^{\dag\mu} \otimes V,\qquad \mbox{for} ~Z_{2m+1},\\
X_{/\!\!/}^{\mu}
&=&H_1^{\mu}\otimes 1+ A_{1}^{\mu}\otimes V^{\dag}+ \cdots + 
A_{m-1}^{\mu}\otimes (V^{\dag})^{2m-3}+ H_{m}^{\mu}\otimes (V^{\dag})^{2m-2} 
\nonumber\\
&&+ H_{m+1}^{\mu}\otimes (V^{\dag})^{2m-1}+ H_{m}^{\dag\mu}\otimes V^{2m-2} + 
A_{m-1}^{\dag\mu}\otimes V^{2m-3}+ \cdots \nonumber\\
&&+ H_{2}^{\mu}\otimes V^2 + A_{1}^{\dag\mu}\otimes V, 
\qquad\qquad\qquad\qquad\qquad\mbox{for}~Z_{4m-2},\\
X_{/\!\!/}^{\mu}&=&
H_1^{\mu}\otimes 1+ A_{1}^{\mu}\otimes V^{\dag}+ \cdots + 
H_m^{\mu}\otimes (V^{\dag})^{2m-2} + A_{m}^{\mu}\otimes (V^{\dag})^{2m-1} 
\nonumber\\
&&+ H_{m+1}^{\mu}\otimes (V^{\dag})^{2m}+ A_{m}^{\dag\mu}\otimes V^{2m-1} +
H_m^{\mu}\otimes V^{2m-2}+ \cdots \nonumber\\
&&+ H_2^{\mu}\otimes V^2 + A_1^{\mu}\otimes V,
\qquad\qquad\qquad\qquad\qquad\mbox{for}~Z_{4m},
\end{eqnarray}
\end{subequations}
\begin{eqnarray}
Z_i&=&
B_{1i}\otimes U+B_{2i} \otimes UV^{\dag} + B_{3i} \otimes U(V^{\dag})^2 +
\cdots + B_{(n-1)i}\otimes UV^2 \nonumber\\
&&+ B_{ni}\otimes UV,
\end{eqnarray}
\begin{subequations}
\begin{eqnarray}
\Theta_1&=&
{\hat H}_1\otimes 1+{\hat A}_1\otimes V^{\dag}+ 
{\hat A}_{2}\otimes (V^{\dag})^{2}+
\cdots +{\hat A}_{m}\otimes (V^{\dag})^{m}\nonumber\\
&&+{\hat A}_{m}^{\dag}\otimes V^{m}+\cdots + 
{\hat A}_{2}^{\dag}\otimes V^2 +
{\hat A}_{1}^{\dag} \otimes V,\qquad \mbox{for} ~Z_{2m+1},\\
\Theta_1&=&
{\hat H}_1\otimes 1+ {\hat A}_{1}\otimes V^{\dag}+ \cdots + 
{\hat A}_{m-1}\otimes (V^{\dag})^{2m-3}+ 
{\hat H}_{m}\otimes (V^{\dag})^{2m-2}\nonumber\\
&&+ {\hat H}_{m+1}\otimes (V^{\dag})^{2m-1}
+ {\hat H}_{m}^{\dag}\otimes V^{2m-2} + 
{\hat A}_{m-1}^{\dag}\otimes V^{2m-3}+ \cdots \nonumber\\
&&+ {\hat H}_{2}\otimes V^2 + 
{\hat A}_{1}^{\dag}\otimes V, \qquad\qquad\qquad\qquad\qquad
\mbox{for}~Z_{4m-2},\\
\Theta_1&=&
{\hat H}_1\otimes 1+ {\hat A}_{1}\otimes V^{\dag}+ \cdots + 
{\hat H}_m\otimes (V^{\dag})^{2m-2} + 
{\hat A}_{m}\otimes (V^{\dag})^{2m-1} \nonumber\\
&&+ {\hat H}_{m+1}\otimes (V^{\dag})^{2m}
+ {\hat A}_{m}^{\dag}\otimes V^{2m-1} +
{\hat H}_m\otimes V^{2m-2}+ \cdots \nonumber\\
&&+ {\hat H}_2\otimes V^2 + 
{\hat A}_1\otimes V, \qquad\qquad\qquad\qquad\qquad\mbox{for}~Z_{4m},
\end{eqnarray}
\end{subequations}
\begin{eqnarray}
\Theta_{\omega}
&=& {\hat B}_1\otimes U+{\hat B}_2 \otimes UV^{\dag} + 
{\hat B}_3 \otimes U(V^{\dag})^2 +
\cdots + {\hat B}_{n-1}\otimes UV^2 \nonumber\\
&&+ {\hat B}_{n}\otimes UV,
\end{eqnarray}
where $H_i$ and ${\hat H}_i$ are $N\times N$ hermitian matrices 
and $A_i$, $B_i$, ${\hat A}_i$ and ${\hat B}_i$ are arbitrary matrices.
The degrees of freedom of $N\times N$ hermitian matrix is equal to
$N^2$, and the degrees of freedom of $N\times N$ arbitrary matrix is
equal to $2N^2$. 
Therefore we can obtain the degrees of freedom of 
the bosonic fields and the fermionic fields for eq. 
(\ref{special C^3/Z_{2,3,6}, C^2/Z_{2,4}, C^1/Z_2}).

%%%%%%%% Figure  %%%%%%%%%%%%%%
\begin{table}[t,h,b]
\caption[The degrees of freedom on $C^i/Z_n$~ $(i=2,3,~n=2,3,4,6)$]
{The degrees of freedom on $C^i/Z_n$~ $(i=2,3,~n=2,3,4,6)$}
\label{table.C^i/Z_n}
\begin{center}
\begin{tabular}{|l|l|l|l|l|}
\hline
$H/\Gamma$ & $X_{/\!\!/}^{\mu}$ & $Z_{i}$ & $\Theta_1$ & 
$\Theta_{\omega}$ \\
\hline
$C^3/Z_{2m+1}$ & $2(2m+1)N^2$ & $12(2m+1)N^2$ & 
$4(2m+1)N^2$ & $24(2m+1)N^2$ \\
$C^2/Z_{2m+1}$ & $4(2m+1)N^2$ & $8(2m+1)N^2$ & 
$8(2m+1)N^2$ & $16(2m+1)N^2$ \\
\hline
$C^3/Z_{4m-2}$ & $2(3m-1)N^2$ & $12(4m-2)N^2$ & 
$4(3m-1)N^2$ & $24(4m-2)N^2$ \\
$C^2/Z_{4m-2}$ & $4(3m-1)N^2$ & $8(4m-2)N^2$ & 
$8(3m-1)N^2$ & $16(4m-2)N^2$ \\
\hline
$C^3/Z_{4m}$ & $2(3m+1)N^2$ & $12\cdot 4mN^2$ & 
$4(3m+1)N^2$ & $24\cdot 4mN^2$ \\
$C^2/Z_{4m}$ & $4(3m+1)N^2$ & $8\cdot 4mN^2$ & 
$8(3m+1)N^2$ & $16\cdot 4mN^2$ \\
\hline
\end{tabular}
\end{center}
\end{table}
As we can see in the table \ref{table.C^i/Z_n}, we find that the fermionic
fields are twice of the degrees of freedom of the bosonic fields, 
where $m$ is an integer and $m\geq 1$. 
Especially, we find that the fermionic fields $\Theta_{2\pi i}$ in the 
flat spaces are twice of the degrees of freedom of the bosonic fields
$X_{/\!\!/}^{\mu}$ in the flat spaces and the fermionic fields 
$\Theta_{\omega}$ ($\Theta_{\omega}^{\dag}$) in the orbifolded 
spaces are twice of the degrees 
of freedom of the bosonic fields $Z_j$ ($Z_{j}^{\dag}$).

We find that the result in the table \ref{table.C^i/Z_n} 
may implicitly mean the existence of the supersymmetry. 
Therefore we will actually check whether the 
supersymmetry exists or not in the next subsection.
However we notice that for the degrees of freedom of fermions in the case of
$c=-1$  we find that $\Theta_{2\pi i}^*$ is equal to $\Theta_{2\pi
i/2}$ from eq. 
(\ref{c=pm1 majorana condition}) and in table \ref{fig.c3/z2_group} of 
the appendix. We then find that matrices 
$\Theta_{2\pi i/2}$ have the same degrees of freedom as matrices
$\Theta_{2\pi i}$. Namely we find that degrees of freedom of bosonic 
fields are not same as that of fermionic fields on the $C^3/Z_2$ 
orbifolds using the representation of 
$(\frac{1}{2},\frac{1}{2}, \frac{1}{2})$ and on the
$C^1/Z_2$ orbifolds using the representation of 
$(1,1,\frac{1}{2})$ shown in (\ref{special
C^3/Z_{2,3,6}, C^2/Z_{2,4}, C^1/Z_2}). Therefore we need to check 
the supersymmetry on the $C^2/Z_2$, $C^2/Z_4$ and $C^3/Z_6$ orbifolds
except $C^3/Z_2$ and $C^1/Z_2$ orbifolds of eq. (\ref{special
C^3/Z_{2,3,6}, C^2/Z_{2,4}, C^1/Z_2}). 

\subsection*{SUSY of IIA Matrix model on the special $C^i/Z_n$ Orbifold}
We confirm the existence of the supersymmetry on the $C^2/Z_2$, $C^2/Z_4$
and $C^3/Z_6$ orbifolds in the similar way as we checked for the $C^3/Z_3$ 
orbifold \cite{Miyake-Sugamoto}. 
The supercharge of IIA Matrix model in 10 dimensions is given by
\begin{eqnarray}
{\bar \epsilon}'Q
&=&{\rm Tr}\left\{\frac{1}{g}\left(i\Theta^{\dag}\right)_{\alpha}
\left(\Gamma^{0I}D_t X_I-i
\Gamma^{IJ}[X_I,X_J]\right)\epsilon_{\alpha}+\left
(i\Theta^{\dag}\right)_{\alpha}\xi_{\alpha}\right\},\label{supercharge}
\end{eqnarray}
and 
\begin{eqnarray}
\epsilon'=
\left(\begin{array}{c}
\epsilon^{[i]a}b_0^{\dag}b_i^{\dag}|0\rangle\\
\epsilon^{[ijk]a}b_0^{\dag}b_{i}^{\dag}b_{j}^{\dag}b_{k}^{\dag}|0\rangle
\end{array}\right),
\end{eqnarray}
where $a=1,2$, $i,j,k=1,\cdots ,4$, $\epsilon'\equiv (\epsilon^1,
\epsilon^2)^T \equiv (\epsilon,\xi)^T$ 
and $\epsilon'$ are sixteen Majorana-Weyl fermions on the mass shell, 
i.e., $\epsilon^{[i]a*}|0\rangle=\frac{1}{6}
\epsilon_{ijkl}\epsilon^{[jkl]a*}|0\rangle$.
Since there are no the bosonic 
fields in the second term on the right hand side of eq. (\ref{supercharge}), 
we can not verify the supersymmetric transformation.
We then omit the second term  of
eq. (\ref{supercharge}):
\begin{eqnarray}
{\bar \epsilon}Q=\frac{1}{g}{\rm Tr}\left[(i\Theta^{\dag})_{\alpha}\left(
\Gamma^{0I}D_t X_I-i
\Gamma^{IJ}[X_I,X_J]\right)\epsilon_{\alpha}\right].
\end{eqnarray} 

Taking the trace with respect to $n\times n$ matrices, we
obtain the supercharge on the special $C^2/Z_2$, $C^2/Z_4$ and
$C^3/Z_6$ orbifolds. The supercharges are given in eqs. (\ref{supercharge
of c2/z2}), (\ref{supercharge of c2/z4}) and (\ref{supercharge of c3/z6}) 
in the appendix.
From the result of the supercharges on $C^2/Z_2$, $C^2/Z_4$, and
$C^3/Z_6$ orbifolds in eqs. (\ref{supercharge
of c2/z2}), (\ref{supercharge of c2/z4}) and (\ref{supercharge of c3/z6}) 
of the appendix, respectively,
we find that supersymmetry exists. We especially point out that
supersymmetric parameters of the standard Matrix model with respect to
supercharges on $C^2/Z_2$ and
$C^2/Z_4$ orbifolds have half as many supersymmetric parameters as that
of IIA Matrix model on
flat spaces in 10 dimensions. 

SUSY parameters on $C^2/Z_2$ and $C^2/Z_4$ orbifolds are described as
follows:  
\begin{eqnarray}
\left(\begin{array}{c}
\epsilon^{[m]}b_{0}^{\dag}b_{m}^{\dag}|0\rangle\\
\frac{1}{2}\epsilon^{[nij]}b_0^{\dag}b_{n}^{\dag}b_{i}^{\dag}b_{j}^{\dag}
|0\rangle
\end{array}\right),\label{susy parameter on C^/Z_2,4}
\end{eqnarray}
where $m,n=1,2$ and $i,j=3,4$. 
According to eqs. (\ref{supercharge of c2/z2}) and  (\ref{supercharge of
c2/z4}), we find that the supersymmetric parameters $\epsilon^{[m]}$ are 
only used and the remainders $\epsilon^{[i]}$ are not. Then 
${\cal N}=2$ supersymmetry of IIA Matrix model in flat spaces turns into
${\cal N}=1$ supersymmetry of IIA Matrix model on $C^2/Z_2$ and
$C^2/Z_4$ orbifolds. We find that the bosonic fields $X_{/\!\!/}$ are
transformed to fermionic fields $\Theta_{2\pi i}$, by SUSY transformation with
the parameter $\epsilon^{[m]}$, while the bosonic fields $Z_{i}$ are
transformed to the fermionic fields $\Theta_{\omega}$ by the SUSY
transformation with the parameter $\epsilon^{[m34]}$ and $Z_i^{\dag}$
are transformed to $\Theta_{\omega^2}$ with $\epsilon^{[m]}$.

We similarly find that the supercharge on $C^3/Z_6$
orbifold has the same number of supersymmetric parameters as IIA Matrix
model on the flat space
in 10 dimensions. SUSY parameters are described as follows:
\begin{eqnarray}
\left(\begin{array}{c}
\epsilon^{[1]}b_0^{\dag}b_{1}^{\dag}|0\rangle\\
\epsilon^{[i]}b_0^{\dag}b_{i}^{\dag}|0\rangle\\ 
\epsilon^{[234]}b_0^{\dag}b_{2}^{\dag}b_{3}^{\dag}
b_{4}^{\dag}|0\rangle\\
\frac{1}{2}\epsilon^{[1ij]}b_0^{\dag}b_{i}^{\dag}b_{j}^{\dag}
|0\rangle
\end{array}\right),\label{susy parameter on C^3/Z_6}
\end{eqnarray}
where $i=2,\cdots ,4$.
The supersymmetric parameters  on $C^3/Z_6$ orbifold are equal to the
supersymmetric parameters on $C^3/Z_3$ orbifold \cite{Miyake-Sugamoto}.
Namely the bosonic fields $X_{/\!\!/}$  are
transformed to fermionic fields $\Theta_1$, by SUSY transformation with
the parameter $\epsilon^{[1]}$, while the bosonic fields $Z_{i}$ are
transformed to the fermionic fields $\Theta_{\omega}$ by the SUSY
transformation with the parameter $\epsilon^{[i]}$ and $Z_i^{\dag}$ are
transformed to $\Theta_{\omega^2}$ with $\epsilon^{[i]*}$.

\section*{IIA Matrix model on a parity-like $Z_2$ Orbifold}
Finally, we mention about another $C^i/Z_2$ orbifold, i.e. a
$parity like$ $C^i/Z_2$
$orbifold$. Due to N. Kim and S. J. Rey, the fields of IIA Matrix model on
$Z_2$ orbifold is described as follows:
\begin{eqnarray}
X_{/\!\!/}^{\mu}&=&MX_{/\!\!/}^{\mu T}M^{\dag},\nonumber\\
X_{\bot}^{i}&=&-MX_{\bot}^{i T}M^{\dag},\nonumber\\
\Theta &=&{\cal P}M\Theta M^{\dag},
\end{eqnarray}
where ${\cal P}$ is a parity transformation matrix and the bosonic and
fermionic fields have $2N\times 2N$ hermitian matrices \cite{S.J.Rey}. The 
bosonic and fermionic fields have $2N\times 2N$ hermitian matrices.
${\cal P}$ has the same peculiarity to the parity-like $Z_2$ orbifold
because the parity transformation operator is equal to the phase of
$Z_2$ symmetry:
\begin{eqnarray}
{\cal P}&\equiv& (i\Gamma^6\Gamma^7)(i\Gamma^8\Gamma^9)\nonumber\\
&=&\exp\left(-2\pi i\sum_{i=3}^{4}\left(-\frac{n_i}{2}\right)
b_i^{\dag}b_{i}\right)={\hat \omega},
\end{eqnarray}
where the matrix $M$ satisfy $M^2=1$ and $M M^T = M^T M=\pm 1$, and so
$M$ belongs to $SO(2N)$ or $USp(2N)$ group. 
A parity-like $Z_2$ orbifold can be defined not only in even dimensions
but also defined in any dimensions.  
In the case of $SO(2N)$ group, the matrix $M$ is defined as
\begin{eqnarray}
M=\left(\begin{array}{cc}
0 & 1 \\
1 & 0
\end{array}\right)_{2N\times 2N}.
\end{eqnarray}
Then the matrices $X_{/\!\!/}^{\mu}$, $X_{\bot}^{i}$, $\Theta_1$ and
$\Theta_{\omega}$ are written as follows:
\begin{eqnarray}
X_{/\!\!/}^{\mu}=\left(\begin{array}{cc}
H_1^\mu & S^\mu \\
S^{*\mu} & H_1^{T\mu}
\end{array}\right),\qquad
X_{\bot}^{i}=\left(\begin{array}{cc}
H_2^i & J^i \\
-J^{*i} & -H_2^{Ti}
\end{array}\right),
\end{eqnarray}
\begin{eqnarray}
\Theta_{2\pi i}=\left(\begin{array}{cc}
{\hat H}_1 & {\hat S} \\
{\hat S}^{*} & {\hat H}_1^T
\end{array}\right),\qquad
\Theta_{\omega}=
\left(\begin{array}{cc}
{\hat H}_2 & {\hat J} \\
-{\hat J}^{*} & -{\hat H}_2^{T}
\end{array}\right),
\end{eqnarray}
where $H_1$, $H_2$, ${\hat H}_1$ and ${\hat H}_2$ are $N\times N$
hermitian matrices, $S$ and ${\hat S}$ are $N\times N$ symmetric
matrices and $J$ and ${\hat J}$ are $N\times N$ anti-symmetric matrices.
While, in the case of $USp(2N)$ group the matrix $M$ is defined as
\begin{eqnarray}
M=\left(\begin{array}{cc}
0 & -i \\
i & 0
\end{array}\right)_{2N\times 2N}.
\end{eqnarray}
Then the matrices $X_{/\!\!/}^{\mu}$, $X_{\bot}^{i}$, $\Theta_{2\pi i}$ and
$\Theta_{\omega}$ are written as
\begin{eqnarray}
X_{/\!\!/}^{\mu}=\left(\begin{array}{cc}
H_1^\mu & J^\mu \\
-J^{*\mu} & H_1^{T\mu}
\end{array}\right),\qquad
X_{\bot}^{i}=\left(\begin{array}{cc}
H_2^i & S^i \\
S^{*i} & -H_2^{Ti}
\end{array}\right),
\end{eqnarray}
\begin{eqnarray}
\Theta_{2\pi i}=\left(\begin{array}{cc}
{\hat H}_1 & {\hat J} \\
-{\hat J}^{*} & {\hat H}_1^{T}
\end{array}\right),\qquad
\Theta_{\omega}=\left(\begin{array}{cc}
{\hat H}_2 & {\hat S} \\
{\hat S}^{*} & -{\hat H}_2^{T}
\end{array}\right).
\end{eqnarray}
From the above fields we can obtain the degrees of freedom of the
bosonic fields and the fermionic fields on the parity-like $Z_2$
orbifold. Then we find that the degrees of freedom of the bosonic 
fields is not equal to that of fermionic fields on $C^1/Z_2$ and
$C^3/Z_2$ orbifolds. On the other hand the degrees of freedom of bosonic 
and fermionic fields are same for $C^2/Z_2$ orbifold. Hereafter we only 
treat $C^2/Z_2$ orbifold. 
\begin{table}[h]
\caption[The degrees of freedom of $SO(2N)$ group on parity-like $Z_2$
 orbifold]{The degrees of freedom of $SO(2N)$ group on parity-like $Z_2$
 orbifold}
\label{fig.SO(2N)}
\begin{center}
\begin{tabular}{|l|l|l|l|l|}
\hline
 & $X_{/\!\!/}^{\mu}$ & $X_{\bot}^{i}$ & $\Theta_{2\pi i}$ &
 $\Theta_{\pi i}$\\
\hline
$R^4/Z_2$ & $2N(2N+1)/2 \times 4$ & $2N(2N-1)/2 \times 4$ & 
$4N(2N+1)/2 \times 8$ & $4N(2N-1)/2 \times 8$ \\
\hline
\end{tabular}
\end{center}
%\end{table}
%\begin{table}[h]
\caption[The degrees of freedom of $USp(2N)$ group on parity-like $Z_2$
 orbifold]{The degrees of freedom of $USp(2N)$ group on parity-like $Z_2$
 orbifold}
\label{fig.USp(2N)}
\begin{center}
\begin{tabular}{|l|l|l|l|l|}
\hline
 & $X_{/\!\!/}^{\mu}$ & $X_{\bot}^{i}$ & $\Theta_{2\pi i}$ &
 $\Theta_{\pi i}$\\
\hline
$R^4/Z_2$ & $2N(2N-1)/2 \times 4$ & $2N(2N+1)/2 \times 4$ & 
$2N(2N-1)/2 \times 8$ & $2N(2N+1)/2 \times 8$ \\
\hline
\end{tabular}
\end{center}
\end{table}

We similarly check the supersymmetry of a parity-like $C^2/Z_2$
orbifold and obtain eq. (\ref{supercharge of parity-like c2/z2}) in
the appendix. SUSY 
parameters on parity-like $C^2/Z_2$ orbifold are described as
\begin{eqnarray}
\left(\begin{array}{c}
\epsilon^{[m]}b_{0}^{\dag}b_{m}^{\dag}|0\rangle\\
\epsilon^{[i]}b_{0}^{\dag}b_{i}^{\dag}|0\rangle\\
\frac{1}{2}\epsilon^{[nij]}b_0^{\dag}b_{n}^{\dag}b_{i}^{\dag}b_{j}^{\dag}
|0\rangle\\
\frac{1}{2}\epsilon^{[mnj]}b_0^{\dag}b_{m}^{\dag}b_{n}^{\dag}b_{j}^{\dag}
|0\rangle
\end{array}\right),\label{susy parameter on parity-like C^2/Z_2}
\end{eqnarray}
where $m,n=1,2$ and $i,j=3,4$. This result on a parity-like $C^2/Z_2$
orbifold is different from that on a normal $C^2/Z_2$ orbifold. Namely
this parity-like $C^2/Z_2$ orbifold has ${\cal N}=2$ supersymmetry.
The bosonic fields $X_{/\!\!/}^{\mu}$  are transformed to the fermionic fields
$\Theta_{2\pi i}$ and $\Theta_{\omega}$, by SUSY transformation with the
parameter $\epsilon^{[m]}$and $\epsilon^{[i]}$, respectively, 
while the bosonic fields $X_{\bot}^{i}$ are transformed to the fermionic
fields $\Theta_{2\pi i}$ and $\Theta_{\omega}$ by the SUSY
transformation with the parameter $\epsilon^{[i]}$ and $\epsilon^{[m]}$,
respectively.

\section*{Conclusions}
In this paper, we have obtained the supersymmetric Matrix models on some
non-compact orbifolds, namely $C^2/Z_2$, $C^2/Z_4$ and $C^3/Z_6$ with a cyclic
identification and  $C^2/Z_2$ with a parity-like identification. 
We considered that the bosonic and fermionic fields consisting of $nN$
D-particles
had the $Z_n$ symmetry, and these fields were represented by
$nN\times nN$ matrices. The fields in 4 dimensions are remained on the
non-orbifolded space-time.
%, and the fields in the rest dimension are imposed
%the $Z_n$ symmetry. 
Imposing the Majonara condition on the orbifolded Matrix model, we
have got eq. (\ref{Majorana condition}) and been able to restrict to the 
special orbifolds, e.g., the $C^2/Z_2$, $C^2/Z_4$, $C^3/Z_3$ and
$C^3/Z_6$ orbifolds. 
The degrees of freedom of
the bosonic and fermionic fields are equal.

Firstly, the number of SUSY parameters of the $C^2/Z_2$ and $C^2/Z_4$
orbifolds with a cyclic identification becomes the half of those of
Matrix models. So the number of the supercharges becomes equal to that 
of ${\cal N}=1$ SUSY in 10 dimensions.

Secondly the number of SUSY parameters of the $C^3/Z_6$ orbifold with a
cyclic identification becomes equal to the case of Matrix models. Then the
number of the supercharges remains the same as that of ${\cal N}=2$ SUSY in 10
dimensions. 

In the case of first and second cases, with respect to the
$C^2/Z_2$, $C^2/Z_4$ and
$C^3/Z_6$ orbifolds with a cyclic identification, the bosonic fields
$X_{/\!\!/}^{\mu}$ without imposing the orbifold on the space-time are
transformed to the fermionic fields $\Theta_{2\pi i}$ without imposing
the orbifold on the space-time by the SUSY parameters. 
While the bosonic fields $Z_{i}$ with orbifolding spaces are
transformed to the fermionic fields
$\Theta_{\omega}$ with imposing the orbifolding spaces by the SUSY
parameters.

Finally
the number of degrees of freedom of the $C^2/Z_2$ orbifold with a parity-like
identification is equal to the case of Matrix models. So the number of
the supercharges also remains those the same as ${\cal N}=2$ SUSY in 10
dimensions. 
Then SUSY of the $C^2/Z_2$ orbifold with a parity-like identification
has SUSY of the $C^2/Z_2$ and $C^2/Z_4$ orbifolds with a cyclic
identification and furthermore the bosonic fields
$X_{/\!\!/}^{\mu}~(Z_i)$ are transformed to the fermionic fields
$\Theta_{\omega}~(\Theta_{2\pi i})$. 

Furthermore compactifying for the $C^2/Z_2$ orbifold spaces, we may find
that ${\cal N}=4$ SUSY in 6 dimensions appears.

%If the compactification of Matrix models on
%the supersymmetric orbifolded spaces is demonstrated, we will be able to
%describe the 4d realistic model. In particular when the model is located 
%on the orbifolded spaces, the space of torus divided by its discrete
%symmetry. 
%
\section*{Acknowledgment}
%%% ともに議論してこの論文を陰ながら支えてくれた菅本教授に大変感謝してお
%%% ります。この論文は著者一人では仕上げる事は不可能であったと感じていま
%%% す。また、河本教授には（急なお願いにも関わらず）論文の仕上げ段階につ
%%% いて議論していただいた事に感謝しております。
I would deeply like to thank Prof. A. Sugamoto for collaborating the
previous work and a part of this work with many useful discussions and 
encouraging greatly in the
course of this work. Thanks are also due to Prof. N. Kawamoto for useful 
suggestions in the completion of this paper and reading the manuscript.

\section*{Appendix}
\subsection*{Gamma matrices}
Gamma matrices which belong to the Clifford algebra are defined as
\begin{eqnarray}
\{\Gamma^{\mu}, \Gamma^{\nu}\}&=&2\eta^{\mu\nu},\cr
\label{gamma anticommutation}
\eta^{\mu\nu}&=&diag.(-,+,\ldots ,+),\cr
\Gamma^D &=& \Gamma^0 \Gamma^1 \ldots \Gamma^{D-1},\cr
\Gamma^{\mu}\Gamma^{\mu\dag}&=&\Gamma^{\mu\dag}\Gamma^{\mu}=1.
\end{eqnarray} 
The gamma matrices in 10 dimensions are given as
\begin{eqnarray}
\left(\begin{array}{c}
\Gamma^{0}\\
\Gamma^{1}
\end{array}\right)
&=& \left(\begin{array}{c}
i\sigma_{2}^{(0)}\\
\sigma_{1}^{(0)}
\end{array}\right)
\otimes  (-\sigma_{3}^{(1)})\otimes  (-\sigma_{3}^{(2)})\otimes 
(-\sigma_{3}^{(3)})\otimes  (-\sigma_{3}^{(4)}),\cr
\left(\begin{array}{c}
\Gamma^{2}\\
\Gamma^{3}
\end{array}\right)
&=& 1^{(0)} \otimes
\left(\begin{array}{c}
\sigma_{1}^{(1)}\\
\sigma_{2}^{(1)}
\end{array}\right)
\otimes  (-\sigma_{3}^{(2)})\otimes  (-\sigma_{3}^{(3)})\otimes  
(-\sigma_{3}^{(4)}),\cr
\left(\begin{array}{c}
\Gamma^{4}\\
\Gamma^{5}
\end{array}\right)
&=& 1^{(0)}\otimes  1^{(1)}\otimes 
\left(\begin{array}{c}
\sigma_{1}^{(2)}\\
\sigma_{2}^{(2)}
\end{array}\right)
\otimes  (-\sigma_{3}^{(3)})\otimes  (-\sigma_{3}^{(4)}),\cr
\left(\begin{array}{c}
\Gamma^{6}\\
\Gamma^{7}
\end{array}\right)
&=& 1^{(0)}\otimes  1^{(1)}\otimes  1^{(2)}\otimes 
\left(\begin{array}{c}
\sigma_{1}^{(3)}\\
\sigma_{2}^{(3)}
\end{array}\right)
\otimes  (-\sigma_{3}^{(4)}),\cr
\left(\begin{array}{c}
\Gamma^{8}\\
\Gamma^{9}
\end{array}\right)
&=& 1^{(0)}\otimes  1^{(1)}\otimes  1^{(2)}\otimes  1^{(3)}\otimes 
\left(\begin{array}{c}
\sigma_{1}^{(4)}\\
\sigma_{2}^{(4)}
\end{array}\right),\label{gamma10}
\end{eqnarray}
where $1^{(\mu)}$ are $2\times 2$ unit matrices and
$\sigma_i~(i=1,2,3)$ are Pauli matrices;
\begin{eqnarray}
\sigma_{1}=
\left(\begin{array}{cc}
0&1\\
1&0
\end{array}\right)~,\quad
\sigma_{2}=
\left(\begin{array}{cc}
0&-i\\
i&0
\end{array}\right)~,\quad
\sigma_{3}=
\left(\begin{array}{cc}
1&0\\
0&-1
\end{array}\right).
\end{eqnarray}
A complex conjugate matrix $B$ is defined by $\Gamma^{\mu*}\equiv
B\Gamma^{\mu}B^{-1}$ and is written as follows:
\begin{eqnarray}
B=c\Gamma^{3}\Gamma^{5}\Gamma^{7}\Gamma^{9}\label{B10},
\end{eqnarray}
where $c$ is a phase which satisfies
$(BB^{\dag}=B^{\dag}B=BB^{*}=B^*B=)|c|^{2}=1$ with Majorana condition:
\begin{eqnarray}
\psi^c &\equiv& \psi , \nonumber\\
\psi &=& B^{-1}\psi^{*}.
\end{eqnarray}

\subsection*{Spinors}
Spinors in 10 dimensions are composed of raising and lowering operators
of gamma matrices. The lowering operators are defined as
\begin{eqnarray}
b_{0}=\frac{1}{2}(\Gamma^{1}-\Gamma^{0})~,\quad
b_{j}=\frac{1}{2}(\Gamma^{2j}-i\Gamma^{2j+1}),
\end{eqnarray}
where $j=1,\cdots ,4$. The eigen state of spinors has all down spin
states and is defined as follows:
\begin{eqnarray}
|0\rangle\equiv |-,-,-,-,-\rangle.\label{ground state}
\end{eqnarray}
Imposing the raising operator $b_{\mu}^{\dag}$ $(\mu=0,\cdots ,4)$ on
eq. (\ref{ground state}), the $\mu$-th down spin state turns into the
excited up spin state. From these states we can give type II 16 
Majorana-Weyl fermions in 10 dimensions, and the $0$-th down spin is
fixed on the mass shell.
\begin{eqnarray}
|\psi \rangle=\left(\begin{array}{c}
\psi |0\rangle\\
\frac{1}{2}\epsilon_{ij}\psi^{[ij]}b_i^{\dag}b_{j}^{\dag}|0\rangle\\
\psi^{[1234]}b_1^{\dag}b_2^{\dag}b_3^{\dag}b_4^{\dag}|0\rangle
\end{array}\right),\label{spinor states}
\end{eqnarray}
where $|\psi\rangle$ are Majorana-Weyl spinors, i.e., 
$\psi^{[1234]*} |0\rangle =\psi |0\rangle$ and
$\psi^{[ij]*}|0\rangle =-\frac{1}{2}\epsilon_{ijkl}\psi^{[kl]}|0\rangle$.

\subsection*{Spinors on the Orbifolds}
We impose the $Z_n$ symmetry for some complex spaces in 10 dimensions
and then we represent spinors corresponding to the complex spaces.
\begin{eqnarray}
{\hat \alpha}|\psi\rangle \equiv \exp\left(2\pi i\sum_{\mu=0}^4 
\frac{n_{\mu}}{n}b_{\mu}^{\dag}b_{\mu}\right)|\psi\rangle,
\label{orbifold spinors}
\end{eqnarray}
where $\mu =0,\cdots ,4$, $n_\mu$ and $n$ are the integer and ${\hat
\alpha}'$ is the operator for $Z_n$ orbifold. 
Imposing Majorana condition, we obtain a condition:
\begin{eqnarray}
\exp\left(2\pi i\sum_{j=1}^{4}\frac{n_j}{n}\right)=c.
\label{orbifold Majorana condition}
\end{eqnarray}
Because of $c^2=1~ (B^2\psi =B\psi^* =\psi)$, we find that $\sum
\frac{n_{j}}{n}$ in eq.
(\ref{orbifold Majorana condition}) has the integer or
half-integer. 
{\renewcommand{\baselinestretch}{1.1}
\begin{table}[thb]
\caption[Spinors on $Z_2$ orbifolds]{Spinors on $Z_2$ orbifolds}
\label{fig.c3/z2_group}
\begin{center}
\begin{tabular}{|l|l|l|l|l|l|l|}
\hline
$(\frac{n_2}{2},\frac{n_3}{2},\frac{n_4}{2})$ & 
$\Theta_{2\pi i}$ & $\Theta_{2\pi i/6}$ & $\Theta_{2\pi i2/6}$ &
 $\Theta_{2\pi i3/6}$ & $\Theta_{2\pi i4/6}$ & $\Theta_{2\pi i5/6}$ \\
\hline
$(\frac{1}{2}, \frac{1}{2}, \frac{2}{2})$ & $2^a, 2^{a*}$ &
 &  & $2^{a'}, 2^{a*'}$ & &  \\
\hline
$(\frac{2}{2},\frac{2}{2},\frac{2}{2})$ & $4^a, 4^{a*}$ &
 & &  &  & \\
\hline
$(\frac{1}{2},\frac{1}{2},\frac{1}{2})$ & $4^a$ & & & 
$4^{^a*}$ & & \\
\hline
$(\frac{1}{2},\frac{2}{2},\frac{2}{2})$ & $4^{a}$ & & & 
$4^{^a*}$ &  &  \\
\hline
\end{tabular}
\end{center}
%\end{table}
%\begin{table}[thb]
\caption[Spinors on $Z_3$ orbifolds]{Spinors on $Z_3$ orbifolds}
\label{fig.c3/z3_group}
\begin{center}
\begin{tabular}{|l|l|l|l|l|l|l|}
\hline
$(\frac{n_2}{3},\frac{n_3}{3},\frac{n_4}{3})$ & 
$\Theta_{2\pi i}$ & $\Theta_{2\pi i/6}$ & $\Theta_{2\pi i2/6}$ &
 $\Theta_{2\pi i3/6}$ & $\Theta_{2\pi i4/6}$ & $\Theta_{2\pi i5/6}$ \\
\hline
$(\frac{1}{3}, \frac{1}{3}, \frac{1}{3})$ & $1^{a}, 1^{a*}$ &
 & $3^{a}$ & & $3^{a*}$ & \\
$(\frac{2}{3},\frac{2}{3},\frac{2}{3})$ & $1^{a}, 1^{a*}$ &
 & $3^{a*}$ & & $3^a$ & \\
\hline
$(\frac{1}{3},\frac{2}{3},\frac{3}{3})$ & $2^{a}, 2^{a*}$ & &
 $2^{a'}$ & & $2^{a*'}$ & \\
\hline
$(\frac{3}{3},\frac{3}{3},\frac{3}{3})$ & $4^{a}, 4^{a*}$ &
 & & & & \\
\hline
\end{tabular}
\end{center}
%\end{table}
%\begin{table}[thb]
\caption[Spinors on $Z_4$ orbifolds]{Spinors on $Z_4$ orbifolds}
\label{fig.c3/z4_group}
\begin{center}
\begin{tabular}{|l|l|l|l|l|}
\hline
$(\frac{n_2}{4},\frac{n_3}{4},\frac{n_4}{4})$ & 
$\Theta_{2\pi i}$ & $\Theta_{2\pi i/4}$ & $\Theta_{2\pi i2/4}$ &
 $\Theta_{2\pi i3/4}$ \\
\hline
$(\frac{1}{4}, \frac{1}{4}, \frac{2}{4})$ & $1^{a},1^{a*}$ &
$2^{a}$ & $1^{a'},1^{a*'}$ & $2^{a*}$ \\
$(\frac{2}{4},\frac{3}{4},\frac{3}{4})$ & $1^{a},1^{a*}$ &
$2^{a*}$ & $1^{a'},1^{a*'}$ & $2^{a}$ \\
\hline
$(\frac{1}{4},\frac{3}{4},\frac{4}{4})$ & $2^{a},2^{a*}$ &
$2^{a'}$ & & $2^{a*'}$  \\
$(\frac{2}{4},\frac{2}{4},\frac{4}{4})$ & $2^{a},2^{a*}$ &
 & $2^{a'},2^{a*'}$ & \\
\hline
$(\frac{4}{4},\frac{4}{4},\frac{4}{4})$ & $4^a,4^{a*}$ &
 &  &  \\
\hline
$(\frac{1}{4},\frac{2}{4},\frac{3}{4})$ & $2^{a}$ & $1^{a},
1^{a*}$ & $2^{a*}$ & $1^{a'},1^{a*'}$ \\
$(\frac{2}{4},\frac{2}{4},\frac{2}{4})$ & $4^{a}$ & & $4^{a*}$
& \\
\hline
$(\frac{1}{4},\frac{1}{4},\frac{4}{4})$ & $2^{a}$ & $2^{a'},
2^{a*'}$ & $2^{a*}$ & \\
$(\frac{3}{4},\frac{3}{4},\frac{4}{4})$ & $2^{a}$ & & 
$2^{a*}$  & $2^{a'},2^{a*'}$  \\
\hline
$(\frac{2}{4},\frac{4}{4},\frac{4}{4})$ & $4^{a}$ & &
$4^{a*}$ & \\
\hline
\end{tabular}
\end{center}
\end{table}
\begin{table}[thb]
\caption[Spinors on $Z_6$ orbifolds]{Spinors on $Z_6$ orbifolds}
\label{fig.c3/z6_group}
\begin{center}
\begin{tabular}{|l|l|l|l|l|l|l|}
\hline
$(\frac{n_2}{6},\frac{n_3}{6},\frac{n_4}{6})$ & $\Theta_{2\pi i}$ & 
$\Theta_{2\pi i/6}$ & $\Theta_{2\pi i2/6}$ &
 $\Theta_{2\pi i3/6}$ & $\Theta_{2\pi i4/6}$ & $\Theta_{2\pi i5/6}$ \\
\hline
$(\frac{1}{6}, \frac{1}{6}, \frac{4}{6})$ & $1^{a},1^{a*}$ &
$2^{a}$ & $1^{a*'}$ & & $1^{a'}$ & $2^{a*}$ \\
$(\frac{1}{6},\frac{2}{6},\frac{3}{6})$ & $1^{a},1^{a*}$ &
$1^{a}$ & $1^{a''}$ & $1^{a'},1^{a*'}$ & $1^{a*''}$ &
 $1^{a*}$ \\
$(\frac{2}{6},\frac{2}{6},\frac{2}{6})$ & $1^{a},1^{a*}$ & &
 $3^{a}$ & & $3^{a*}$ & \\
$(\frac{2}{6},\frac{5}{6},\frac{5}{6})$ & $1^{a},1^{a*}$ &
$2^{a*}$ & $1^{a'}$ & & $1^{a*'}$ & $2^{a}$ \\
$(\frac{3}{6},\frac{4}{6},\frac{5}{6})$ & $1^{a},1^{a*}$ &
$1^{a*}$ & $1^{a*''}$ & $1^{a'},1^{a*'}$ & $1^{a''}$ &
$1^{a}$ \\
$(\frac{4}{6},\frac{4}{6},\frac{4}{6})$ &  $1^{a},1^{a*}$ & &
 $3^{a*}$ & & $3^{a}$ & \\
\hline
$(\frac{1}{6},\frac{5}{6},\frac{6}{6})$ & $2^{a},2^{a*}$ &
$2^{a'}$ & & & & $2^{a*'}$ \\
$(\frac{2}{6},\frac{4}{6},\frac{6}{6})$ & $2^{a},2^{a*}$ & &
 $2^{a'}$ & & $2^{a*'}$ & \\
$(\frac{3}{6},\frac{3}{6},\frac{6}{6})$ & $2^{a},2^{a*}$ & &
 & $2^{a'},2^{a*'}$  &   & \\
\hline
$(\frac{6}{6},\frac{6}{6},\frac{6}{6})$ & $4^{a},4^{a*}$ & &
 &  & & \\
\hline
$(\frac{1}{6},\frac{1}{6},\frac{1}{6})$ & $1^{a}$ & $3^{a}$ &
 $3^{a*}$ & $1^{a*}$ & & \\
$(\frac{1}{6},\frac{3}{6},\frac{5}{6})$ & $2^{a}$ & $1^{a}$ &
 $1^{a*}$ & $2^{a*}$ & $1^{a*'}$ & $1^{a'}$ \\
$(\frac{1}{6},\frac{4}{6},\frac{4}{6})$ & $1^a$ & $1^{a'}$ &
 $1^{a*'}$ & $1^{a*}$ & $2^a$ & $2^{a*}$ \\
$(\frac{2}{6},\frac{2}{6},\frac{5}{6})$ & $1^{a}$ & $2^{a*}$ &
 $2^a$ & $1^{a*}$ & $1^{a*'}$ & $1^{a'}$ \\
$(\frac{2}{6},\frac{3}{6},\frac{4}{6})$ & $2^{a}$ & $1^{a*}$ &
 $1^{a}$ & $2^{a*}$ & $1^{a'}$ & $1^{a*'}$ \\
$(\frac{3}{6},\frac{3}{6},\frac{3}{6})$ & $4^{a}$ & & &
 $4^{a*}$ & & \\
$(\frac{5}{6},\frac{5}{6},\frac{5}{6})$ & $1^{a}$ & & & $1^{a*}$
 & $3^{a*}$ & $3^{a}$ \\
\hline
$(\frac{1}{6},\frac{2}{6},\frac{6}{6})$ & $2^{a}$ & $2^{a'}$ &
 $2^{a*'}$ & $2^{a*}$ & & \\
$(\frac{4}{6},\frac{5}{6},\frac{6}{6})$ & $2^{a}$ & & &
 $2^{a*}$ & $2^{a*'}$ & $2^{a'}$\\
\hline
$(\frac{3}{6},\frac{6}{6},\frac{6}{6})$ & $4^{a}$ & & &
 $4^{a*}$ & & \\
\hline
\end{tabular}
\end{center}
\end{table}}

We substitute eq. (\ref{spinor states}) into eq. (\ref{orbifold
spinors}) and show the states of orbifolded spinors in table 
\ref{fig.c3/z2_group}, \ref{fig.c3/z3_group}, \ref{fig.c3/z4_group},
which are obtained by imposing the symmetry of the  orbifold with
respect to spinors in 10 dimensions, and
\ref{fig.c3/z6_group} in the case of $n=2,3,4$, and then we write the
rotated spinors for the phase $\omega_j(=\exp (2\pi in_j/n))$ as 
$\Theta_{2\pi in_j/n}$ on the mass shell.
From the table \ref{fig.c3/z2_group},
\ref{fig.c3/z3_group}, \ref{fig.c3/z4_group} and \ref{fig.c3/z6_group}, 
we find that spinors have the next relations:
\begin{eqnarray}
\Theta_{(2\pi i-a)i}^* &=&\Theta_{ai},\quad (c=1),\nonumber\\
\Theta_{(\pi -a)i}^{*} &=&\Theta_{ai},\quad (c=-1).
\label{c=pm1 majorana condition}
\end{eqnarray} 
Then we find that $Z_{2m+1}$
orbifolded spaces in the case of $c=-1$
do not 
exist, which can be understood from the left part of eq.  
(\ref{orbifold Majorana condition}). 

\subsection*{Supercharge}
The conjugate fields for the bosonic fields 
$(X^{I})_{ij}$ and the fermionic fields $(\Theta )_{ij}$
in 10 dimensions are written as
\begin{eqnarray}
(\Pi^I)_{ji}&\equiv&\frac{\partial L}{\partial (D_t X^I)_{ij}}
=\frac{1}{g^2}(D_t X_{I})_{ji},\nonumber\\
(\Pi_{\Theta})_{ji}&\equiv&\frac{\partial L}{\partial (D_t \Theta)_{ij}}=
i(\Theta^{\dag})_{ji},
\end{eqnarray}
where $I=1,\cdots ,9$ and $i,j=1,\cdots ,nN$. The subscript $ij$ denotes
a component of a $nN\times nN$ matrix.
Then, the canonical commutation relations of these fields are described
as
\begin{eqnarray}
[(X^I)_{ij},(\Pi^J)_{ji}]&=&i\delta_{il}\delta_{jk}\delta^{IJ},\nonumber\\
\{(\Theta)_{ij},(\Theta^{\dag})_{kl}\}&=&\frac{1}{2}i\delta_{il}\delta_{jk}.
\end{eqnarray}
Therefore, the supersymmetric charge is given as follows:
\begin{eqnarray}
{\bar \epsilon'}Q
&=&{\rm Tr}\left(\Pi_I \delta X^I +2\Pi_{\Theta}\delta\Theta
\right),\label{general supercharge}
\end{eqnarray}
where $\Theta$ are Majorana-Weyl fermions. 
We find that supercharge of IIA Matrix model in 10 dimensions
is written as eq. (\ref{supercharge}).

\subsection*{Supercharge on orbifolds}
Supercharge on $C^2/Z_2$ orbifold is written as follows:
\begin{eqnarray}
{\bar \epsilon}Q&=&-\frac{2}{g}{\rm Tr}\left[
i\langle\psi|{\hat H}_1{\dot B}_{1m}+{\hat H}_2{\dot B}_{2m}|\epsilon^{[m]}
\rangle\delta_{km}\right.\nonumber\\
&&+\left.i\langle\psi^{[12]}|{\hat H}_1{\dot B}_{1m}+{\hat H}_2{\dot B}_{2m}
|\epsilon^{[m]}\rangle\epsilon_{km}\right.\nonumber\\
&&\left.+i\langle\psi^{[mi]}|{\hat B}_1^{\dag}{\dot B}_{1j}^{\dag}
+{\hat B}_2^{\dag}{\dot B}_{2j}^{\dag}|
\epsilon^{[n]}\rangle\epsilon_{ij}\delta_{mn}\right.\nonumber\\
&&\left.+\langle\psi|{\hat H}_1\left([H_{10},{B}_{1m}]
+[H_{20},{B}_{2m}]\right)|
\epsilon^{[n]}\rangle\delta_{mn}\right.\nonumber\\
&&\left.+\langle\psi^{[12]}|{\hat H}_1\left([H_{10},{B}_{1m}]
+[H_{20},{B}_{2m}]
\right)|\epsilon^{[n]}\rangle\epsilon_{mn}\right.\nonumber\\
&&\left.+\langle\psi|{\hat H}_1\left([H_{10},{B}_{2m}]
+[H_{20},{B}_{1m}]\right)|
\epsilon^{[n]}\rangle\delta_{mn}\right.\nonumber\\
&&\left.+\langle\psi^{[12]}|{\hat H}_1\left([H_{10},{B}_{1m}]
+[H_{20},{B}_{2m}]
\right)|\epsilon^{[n]}\rangle\epsilon_{mn}\right.\nonumber\\
&&\left.+\langle\psi^{[mi]}|{\hat B}_1^{\dag}\left([H_{10},
{B}_{1j}^{\dag}]+[H_{10},{B}_{2j}^{\dag}]\right)
|\epsilon^{[n]}\rangle\delta_{mn}\epsilon_{ij}(-1)
\right.\nonumber\\
&&\left.+\langle\psi^{[mi]}|{\hat B}_2^{\dag}\left\{(\omega H_{20}
{B}_{1j}^{\dag}-{B}_{1j}^{\dag}H_{20})\right.\right.\nonumber\\*
&&\qquad\qquad\quad\left.\left.+(\omega H_{20}
{B}_{2j}^{\dag}-{B}_{2j}^{\dag}H_{20})\right\}|\epsilon^{[n]}
\rangle\delta_{mn}\epsilon_{ij}(-1
)\right.\nonumber\\
&&\left. +h.c. \right],\label{supercharge of c2/z2}
\end{eqnarray}
where $m,n=1,2$, $i,j=3,4$, and we define $B_{1m}$ as
$H_{1(2m)}+iH_{1(2m+1)}$, $B_{2m}$ as $H_{2(2m)}+iH_{2(2m+1)}$ and
${\dot B}_{1m}$ as $\partial_t B_{1m}$ and so on. We find
that the bosonic fields are transformed to the fermionic fields with
supersymmetric parameters $\epsilon^{[n]}$ on $C^i/Z_2$ orbifold and the 
number of the parameters $\epsilon^{[n]}$ is the half of the number of the
parameters $(\epsilon^{[n]}, \epsilon^{[i]})$ for IIA Matrix model.

Supercharge on $C^2/Z_4$ orbifold is written as
\begin{eqnarray}
{\bar \epsilon}Q&=&-\frac{4}{g}{\rm Tr}\left[
i\langle\psi|{\hat H}_1{\dot B}_{1m}+{\hat A}_{1}{\dot B}_{4m}
+{\hat H}_2{\dot B}_{3m}
+{\hat A}_1^{\dag}A_{10}|\epsilon^{[n]}\rangle\delta_{mn}\right.\nonumber\\
&&\left.+i\langle\psi^{[12]}|{\hat H}_1{\dot B}_{1m}+{\hat A}_{1}
{\dot B}_{4m}
+{\hat H}_2{\dot B}_{3m}+{\hat A}_1^{\dag}{\dot B}_{2m}|\epsilon^{[n]}
\rangle\epsilon_{mn}\right.\nonumber\\
&&\left.+i\langle\psi^{[mi]}|{\hat B}_1^{\dag}{\dot B}_{1i}+{\hat B}_2^{\dag}
{\dot B}_{2i}
+{\hat B}_3^{\dag}{\dot B}_{3i}+{\hat B}_4^{\dag}{\dot B}_{4i}
|\epsilon^{[n34]}\rangle\delta_{mn}\epsilon_{ik}\right.\nonumber\\
&&\left.+\langle\psi|{\hat H}_1\left([H_{10},{B}_{1m}]+[A_{10},
{B}_{4m}]+[H_{20},{B}_{3m}]
+[A_{10}^{\dag},{B}_{2m}]\right)|\epsilon^{[n]}\rangle
\delta_{mn}\right.\nonumber\\
&&\left.+\langle\psi^{[12]}|{\hat H}_1\left([H_{10},{B}_{1m}]
+[A_{10},{B}_{4m}]+[H_{20},{B}_{3m}]+[A_{10}^{\dag},{B}_{2m}]
\right)|\epsilon^{[n]}\rangle\epsilon_{mn}\right.\nonumber\\
&&\left.+\langle\psi|{\hat A}_1^{\dag}\left([H_{10},{B}_{2m}]
+[A_{10},{B}_{1m}]
+[H_{20},{B}_{4m}]+[A_{10}^{\dag},{B}_{3m}]\right)|\epsilon^{[n]}
\rangle\delta_{mn}\right.\nonumber\\
&&\left.+\langle\psi^{[12]}|{\hat A}_1^{\dag}\left([H_{10},B_{2m}]
+[A_{10},B_{1m}]+[H_{20},B_{4m}]+[A_{10}^{\dag},B_{3m}]\right)
|\epsilon^{[n]}\rangle\epsilon_{mn}\right.\nonumber\\
&&\left.+\langle\psi|{\hat H}_2\left([H_{10},B_{3m}]+[A_{10},B_{2m}]
+[H_{20},B_{1m}]+[A_{10}^{\dag},B_{4m}]\right)|\epsilon^{[n]}\rangle
\delta_{mn}\right.\nonumber\\
&&\left.+\langle\psi^{[12]}|{\hat H}_2\left([H_{10},B_{3m}]+[A_{10},B_{2m}]
+[H_{20},B_{1m}]+[A_{10}^{\dag},B_{4m}]\right)|\epsilon^{[n]}\rangle
\epsilon_{mn}\right.\nonumber\\
&&\left.+\langle\psi|{\hat A}_1\left([H_{10},B_{4m}]+[A_{10},B_{3m}]
+[H_{20},B_{2m}]+[A_{10}^{\dag},B_{1m}]\right)|\epsilon^{[n]}\rangle
\delta_{mn}\right.\nonumber\\
&&\left.+\langle\psi^{[12]}|{\hat A}_1\left([H_{10},B_{4m}]
+[A_{10},B_{3m}]
+[H_{20},B_{2m}]+[A_{10}^{\dag},B_{1m}]\right)|\epsilon^{[n]}\rangle
\epsilon_{mn}\right.\nonumber\\
&&\left.+\langle\psi^{[mi]}|{\hat B}_1\left([H_{10},B_{1j}^{\dag}]+
(A_{10}B_{2j}^{\dag}-\omega B_{2j}^{\dag}A_{10})+(H_{20}B_{3j}^{\dag}-
\omega^2 B_{3j}^{\dag}H_{20})\right.\right.\nonumber\\
&&\left.\left.\qquad\qquad\qquad +(A_{10}^{\dag}B_{4j}^{\dag}-\omega^{\dag}
B_{4j}^{\dag}A_{10}^{\dag})\right)|\epsilon^{[n]}\rangle\delta_{ij}
\delta_{mn}(-1)\right.\nonumber\\
&&\left.+\langle\psi^{[mi]}|{\hat B}_2\left([H_{10},B_{2j}^{\dag}]+
(A_{10}B_{3j}^{\dag}-\omega B_{3j}^{\dag}A_{10})+(H_{20}B_{4j}^{\dag}-
\omega^2 B_{4j}^{\dag}H_{20})\right.\right.\nonumber\\
&&\left.\left.\qquad\qquad\qquad +(A_{10}^{\dag}B_{1j}^{\dag}-\omega^{\dag}
B_{1j}^{\dag}A_{10}^{\dag})\right)|\epsilon^{[n]}\rangle\delta_{ij}
\delta_{mn}(-1)\right.\nonumber\\
&&\left.+\langle\psi^{[mi]}|{\hat B}_3\left([H_{10},B_{3j}^{\dag}]+
(A_{10}B_{4j}^{\dag}-\omega B_{4j}^{\dag}A_{10})+(H_{20}B_{1j}^{\dag}-
\omega^2 B_{1j}^{\dag}H_{20})\right.\right.\nonumber\\
&&\left.\left.\qquad\qquad\qquad +(A_{10}^{\dag}B_{2j}^{\dag}-\omega^{\dag}
B_{2j}^{\dag}A_{10}^{\dag})\right)|\epsilon^{[n]}\rangle\delta_{ij}
\delta_{mn}(-1)\right.\nonumber\\
&&\left.+\langle\psi^{[mi]}|{\hat B}_4\left([H_{10},B_{4j}^{\dag}]+
(A_{10}B_{1j}^{\dag}-\omega B_{1j}^{\dag}A_{10})+(H_{20}B_{2j}^{\dag}-
\omega^2 B_{2j}^{\dag}H_{20})\right.\right.\nonumber\\*
&&\left.\left.\qquad\qquad\qquad +(A_{10}^{\dag}B_{3j}^{\dag}-\omega^{\dag}
B_{3j}^{\dag}A_{10}^{\dag})\right)|\epsilon^{[n]}\rangle\delta_{ij}
\delta_{mn}(-1)\right.\nonumber\\
&&+\left. h.c.\right],\label{supercharge of c2/z4}
\end{eqnarray}
where $m,n=1,2$, $i,j=3,4$, and we define $B_{1m}$ as
$H_{1(2m)}+iH_{1(2m+1)}$, $B_{2m}$ as $A_{1(2m)}+iA_{1(2m+1)}$, $B_{3m}$ 
as $H_{2(2m)}+iH_{2(2m+1)}$ and $B_{4m}$ as
$A_{1(2m)}^{\dag}+iA_{1(2m+1)}^{\dag}$. 

Supercharge on $C^3/Z_6$ orbifold is written as
\begin{eqnarray}
{\bar \epsilon}Q&=&-\frac{6}{g}{\rm Tr}\left[
i\langle\psi|{\hat H}_1{\dot B}_{11}+{\hat A}_1^{\dag}{\dot B}_{21}
+{\hat H}_2{\dot B}_{31}+{\hat H}_3 {\dot B}_{41}+{\hat H}_2 {\dot B}_{51}
+{\hat A}_1 {\dot B}_{61}|\epsilon^{[1]}
\rangle\right.\nonumber\\
&&\left.+\langle\psi^{[ij]}|{\hat B}_1{\dot B}_{1i}^{\dag}+{\hat B}_2
{\dot B}_{2i}^{\dag}+{\hat B}_3{\dot B}_{3i}^{\dag}+{\hat B}_4
{\dot B}_{4i}^{\dag}+{\hat B}_5{\dot B}_{5i}^{\dag}
+{\hat B}_6{\dot B}_{6i}^{\dag}|\epsilon^{[l]}\rangle
\right.\nonumber\\*
&&\left.\quad\times(\delta_{ik}\delta_{jl}-\delta_{il}\delta_{jk})
\right.\nonumber\\
&&\left.+\langle\psi|{\hat H}_1\left([H_{10},B_{11}]+[A_{10},B_{61}]
+[H_{20},B_{51}]+[H_{30},B_{41}]\right.\right.\nonumber\\
&&\qquad\qquad\quad\left.\left.+[H_{20},B_{31}]+[A_{10}^{\dag},B_{21}]
\right)|\epsilon^{[1]}\rangle\right.\nonumber\\
&&\left.+\langle\psi|{\hat A}_1^{\dag}\left([H_{10},B_{21}]+[A_{10},B_{11}]
+[H_{20},B_{61}]+[H_{30},B_{51}]\right.\right.\nonumber\\
&&\qquad\qquad\quad\left.\left.+[H_{20},B_{41}]
+[A_{10}^{\dag},B_{31}]\right)|\epsilon^{[1]}\rangle\right.\nonumber\\
&&\left.+\langle\psi|{\hat H}_2\left([H_{10},B_{31}]+[A_{10},B_{21}]
+[H_{20},B_{11}]+[H_{30},B_{61}]\right.\right.\nonumber\\
&&\qquad\qquad\quad\left.\left.+[H_{20},B_{51}]
+[A_{10}^{\dag},B_{41}]\right)
|\epsilon^{[1]}\rangle\right.\nonumber\\
&&\left.+\langle\psi|{\hat H}_3\left([H_{10},B_{41}]+[A_{10},B_{31}]
+[H_{20},B_{21}]+[H_{30},B_{11}]\right.\right.\nonumber\\
&&\qquad\qquad\quad\left.\left.+[H_{20},B_{61}]
+[A_{10}^{\dag},B_{51}]\right)
|\epsilon^{[1]}\rangle\right.\nonumber\\
&&\left.+\langle\psi|{\hat H}_2\left([H_{10},B_{51}]+[A_{10},B_{41}]
+[H_{20},B_{31}]+[H_{30},B_{21}]\right.\right.\nonumber\\
&&\qquad\qquad\quad\left.\left.+[H_{20},B_{11}]
+[A_{10}^{\dag},B_{61}]\right)
|\epsilon^{[1]}\rangle\right.\nonumber\\
&&\left.+\langle\psi|{\hat A}_1\left([H_{10},B_{61}]+[A_{10},B_{51}]
+[H_{20},B_{41}]+[H_{30},B_{31}]\right.\right.\nonumber\\
&&\qquad\qquad\quad\left.\left.+[H_{20},B_{21}]
+[A_{10}^{\dag},B_{11}]\right)
|\epsilon^{[1]}\rangle\right.\nonumber\\
&&\left.+\langle\psi^{[ij]}|{\hat B}_1\left([H_{10},B_{1k}^{\dag}]+(A_{10}
B_{2k}^{\dag}-\omega B_{2k}^{\dag}A_{10})+(H_{20}B_{3k}^{\dag}-
\omega^2 B_{3k}^{\dag}H_{20})\right.\right.\nonumber\\
&&\qquad\qquad\quad\left.\left.+(H_{30}B_{4k}^{\dag}-\omega^3 B_{4k}^{\dag}
H_{30})+(H_{20}B_{5k}^{\dag}-\omega^{\dag 2} B_{5k}^{\dag}H_{20})
\right.\right.\nonumber\\
&&\qquad\qquad\quad\left.\left.+(A_{10}^{\dag}B_{6k}^{\dag}-\omega^{\dag}
B_{6k}^{\dag}A_{10}^{\dag})\right)
|\epsilon^{[l]}\rangle(\delta_{jl}\delta_{ik}-\delta_{jk}\delta_{il})\right.
\nonumber\\
&&\left.+\langle\psi^{[ij]}|{\hat B}_2\left([H_{10},B_{2k}^{\dag}]+(A_{10}
B_{3k}^{\dag}-\omega B_{3k}^{\dag}A_{10})+(H_{20}B_{4k}^{\dag}-
\omega^2 B_{4k}^{\dag}H_{20})\right.\right.\nonumber\\
&&\qquad\qquad\quad\left.\left.+(H_{30}B_{5k}^{\dag}-\omega^3 B_{5k}^{\dag}
H_{30})+(H_{20}B_{6k}^{\dag}-\omega^{\dag 2} B_{6k}^{\dag}H_{20})
\right.\right.\nonumber\\
&&\qquad\qquad\quad\left.\left.+(A_{10}^{\dag}B_{1k}^{\dag}-\omega^{\dag}
B_{1k}^{\dag}A_{10}^{\dag})\right)
|\epsilon^{[l]}\rangle(\delta_{jl}\delta_{ik}-\delta_{jk}\delta_{il})\right.
\nonumber\\
&&\left.+\langle\psi^{[ij]}|{\hat B}_3\left([H_{10},B_{3k}^{\dag}]+(A_{10}
B_{4k}^{\dag}-\omega B_{4k}^{\dag}A_{10})+(H_{20}B_{5k}^{\dag}-
\omega^2 B_{5k}^{\dag}H_{20})\right.\right.\nonumber\\
&&\qquad\qquad\quad\left.\left.+(H_{30}B_{6k}^{\dag}-\omega^3 B_{6k}^{\dag}
H_{30})+(H_{20}B_{1k}^{\dag}-\omega^{\dag 2} B_{1k}^{\dag}H_{20})
\right.\right.\nonumber\\
&&\qquad\qquad\quad\left.\left.+(A_{10}^{\dag}B_{2k}^{\dag}-\omega^{\dag}
B_{2k}^{\dag}A_{10}^{\dag})\right)
|\epsilon^{[l]}\rangle(\delta_{jl}\delta_{ik}-\delta_{jk}\delta_{il})\right.
\nonumber\\
&&\left.+\langle\psi^{[ij]}|{\hat B}_4\left([H_{10},B_{4k}^{\dag}]+(A_{10}
B_{5k}^{\dag}-\omega B_{5k}^{\dag}A_{10})+(H_{20}B_{6k}^{\dag}-
\omega^2 B_{6k}^{\dag}H_{20})\right.\right.\nonumber\\
&&\qquad\qquad\quad\left.\left.+(H_{30}B_{1k}^{\dag}-\omega^3 B_{1k}^{\dag}
H_{30})+(H_{20}B_{2k}^{\dag}-\omega^{\dag 2} B_{2k}^{\dag}H_{10})
\right.\right.\nonumber\\
&&\qquad\qquad\quad\left.\left.+(A_{10}^{\dag}B_{3k}^{\dag}-\omega^{\dag}
B_{3k}^{\dag}A_{10}^{\dag})\right)
|\epsilon^{[l]}\rangle(\delta_{jl}\delta_{ik}-\delta_{jk}\delta_{il})\right.
\nonumber\\
&&\left.+\langle\psi^{[ij]}|{\hat B}_5\left([H_{10},B_{5k}^{\dag}]+(A_{10}
B_{6k}^{\dag}-\omega B_{6k}^{\dag}A_{10})+(H_{20}B_{1k}^{\dag}-
\omega^2 B_{1k}^{\dag}H_{20})\right.\right.\nonumber\\*
&&\qquad\qquad\quad\left.\left.+(H_{30}B_{2k}^{\dag}-\omega^3 B_{2k}^{\dag}
H_{30})+(H_{20}B_{3k}^{\dag}-\omega^{\dag 2} B_{3k}^{\dag}H_{10})
\right.\right.\nonumber\\*
&&\qquad\qquad\quad\left.\left.+(A_{10}^{\dag}B_{4k}^{\dag}-\omega^{\dag}
B_{4k}^{\dag}A_{10}^{\dag})\right)
|\epsilon^{[l]}\rangle(\delta_{jl}\delta_{ik}-\delta_{jk}\delta_{il})\right.
\nonumber\\
&&\left.+\langle\psi^{[ij]}|{\hat B}_6\left([H_{10},B_{6k}^{\dag}]+(A_{10}
B_{1k}^{\dag}-\omega B_{1k}^{\dag}A_{10})+(H_{20}B_{2k}^{\dag}-
\omega^2 B_{2k}^{\dag}H_{20})\right.\right.\nonumber\\*
&&\qquad\qquad\quad\left.\left.+(H_{30}B_{3k}^{\dag}-\omega^3 B_{3k}^{\dag}
H_{30})+(H_{20}B_{4k}^{\dag}-\omega^{\dag 2} B_{4k}^{\dag}H_{10})
\right.\right.\nonumber\\*
&&\qquad\qquad\quad\left.\left.+(A_{10}^{\dag}B_{5k}^{\dag}-\omega^{\dag}
B_{5k}^{\dag}A_{10}^{\dag})\right)
|\epsilon^{[l]}\rangle(\delta_{jl}\delta_{ik}-\delta_{jk}\delta_{il})\right.
\nonumber\\
&&\left.+h.c. \right],\label{supercharge of c3/z6}
\end{eqnarray}
where $i,j,k,l=1,2,3$, and we define $B_{11}$ as
$H_{1(2m)}+iH_{1(2m+1)}$, $B_{21}$ as $A_{1(2m)}+iA_{1(2m+1)}$, $B_{31}$ 
as $H_{2(2m)}+iH_{2(2m+1)}$, $B_{41}$ as $H_{3(2m)}+iH_{3(2m+1)}$,
$B_{51}$ as $H_{2(2m)}+iH_{2(2m+1)}$ and $B_{61}$ as $A_{1(2m)}^{\dag}
+iA_{1(2m+1)}^{\dag}$.

Supercharge on a parity like $C^2/Z_2$ orbifold included in $USp(2N)$
symmetry is written as
\begin{eqnarray}
{\bar \epsilon}Q&=&-\frac{1}{g}{\rm Tr}\left[
2i\langle \psi|{\hat H}_1{\dot B}_{1m}-\frac{1}{2}\left({\hat J}
{\dot B}_{2m}^* +{\hat J}^*{\dot B}_{2m}\right)
|\epsilon^{[n]}\rangle\delta_{mn}\right.\nonumber\\
&&+2i\left.\langle\psi^{[12]}|{\hat H}_1{\dot B}_{1m}-\frac{1}{2}
\left({\hat J}{\dot B}_{2m}^* +{\hat J}^*{\dot B}_{2m}\right)
|\epsilon^{[n]}\rangle\epsilon_{mn}
\right.\nonumber\\
&&+2i\left.\langle\psi^{[mi]}|{\hat H}_2{\dot B}_{1j}+\frac{1}{2}\left(
{\hat S}{\dot B}_{2j}^*+{\hat S}^* {\dot B}_{2j}\right)
|\epsilon^{[n]}\rangle 
\delta_{ij}\delta_{mn}(-1)\right.\nonumber\\
&&+\left.\langle\psi|{\hat H}_1\left\{(H_{10}B_{1m}-J_{0}{B}_{2m}^*)-
(B_{1m}B_{10}-B_{2m}J_{0}^*)\right\}\right.\nonumber\\
&&\qquad\left.+{\hat J}\left\{(-J_{0}^*B_{1m}-H_{10}^TB_{2m}^*)-
(-B_{2m}^*H_{10}-B_{1m}J_{\mu}^*)\right\}\right.\nonumber\\
&&\qquad\left.-{\hat J}^*\left\{(H_{10}B_{2m}+J_{0}B_{1m})-
(B_{1m}J_{0}-B_{2m}J_{0}^T)\right\}\right.\nonumber\\
&&\qquad\left.+{\hat H}_1^{T}\left\{(-J_{0}^*B_{2m}+H_{10}^TB_{1m})-
(-B_{2m}^*J_{0}+B_{1m}H_{10}^{T})\right\}|\epsilon^{[n]}\rangle\delta_{mn}
\right.\nonumber\\
&&+\left.\langle\psi^{[12]}|{\hat H}_1\left\{(H_{10}B_{1m}-J_{0}B_{2m}^*)-
(B_{1m}H_{10}-B_{2m}J_{0}^*)\right\}\right.\nonumber\\
&&\qquad\left.+{\hat J}\left\{(-J_{0}^*B_{1m}-H_{10}^TB_{2m}^*)-
(-B_{2m}^*H_{10}-B_{1m}J_{0}^*)\right\}\right.\nonumber\\
&&\qquad\left.-{\hat J}^*\left\{(H_{10}B_{2m}+J_{0}B_{1m})-
(B_{1m}J_0-B_{2m}H_{10}^T)\right\}\right.\nonumber\\
&&\qquad\left.+{\hat H}_1^{T}\left\{(-J_{0}^*B_{2m}+H_{10}^TB_{1m})-
(-B_{2m}^*J_{0}+B_{1m}H_{10}^{T})\right\}|\epsilon^{[n]}\rangle\epsilon_{mn}
\right.\nonumber\\
&&\left.+\langle\psi|{\hat H}_1\left\{(H_{10}B_{1i}+J_{0}B_{2i}^*)-
(H_{2i}H_{1\mu}-B_{2j}J_0^*)\right\}|\epsilon^{[j]}\rangle\delta_{ij}
\right.\nonumber\\
&&\left.+\langle\psi|{\hat J}\left\{(-J_{0}^*B_{2m}+H_{10}^TB_{1m})-
(-B_{2m}^*J_0+B_{1m}H_{10}^T)\right\}\right.\nonumber\\
&&\qquad\left.-{\hat J}^*\left\{(H_{10}B_{2m}+J_{0}B_{1m})-
(B_{1m}^*J_{0}+B_{2m}H_{10}^T)\right\}|\epsilon^{[n]}\rangle\delta_{mn}
\right.\nonumber\\
&&\left.+\langle\psi|{\hat H}_1^{T}\left\{(-B_{20}^*B_{2i}
-B_{10}^TB_{1i}^T)-(B_{2i}^*B_{20}-B_{1i}^TB_{10})
\right\}|\epsilon^{[j]}\rangle\delta_{ij}\right.\nonumber\\
&&+\left.\langle\psi^{[12]*}|{\hat H}_1\left\{(H_{10}B_{1i}
+J_{0}B_{1i}^*)-(B_{1i}H_{10}-B_{2i}J_0^*)\right\}|\epsilon^{[j]}
\rangle\epsilon_{ij}\right.\nonumber\\
&&\left.+\langle\psi^{[12]}|{\hat J}\left\{(-J_{0}^*B_{2m}+H_{10}^T
B_{1m})-(-B_{2m}^*J_0+B_{1m}H_{10}^T)\right\}\right.\nonumber\\
&&\qquad\left.-{\hat J}^*\left\{(H_{10}B_{2m}+J_{0}B_{1m})-
(B_{1m}^*J_{0}+B_{2m}H_{10}^T)\right\}|\epsilon^{[n]}\rangle
\epsilon_{mn}\right.\nonumber\\
&&\left.+\langle\psi^{[12]*}|{\hat H}_1^{T}\left\{(-B_{20}^*B_{2i}
-B_{10}^TB_{1i}^T)-(B_{2i}^*B_{20}-B_{1i}^TB_{10})
\right\}|\epsilon^{[j]}\rangle\epsilon_{ij}\right.\nonumber\\
&&+\left.\langle\psi^{[mi]}|{\hat H}_2\left\{(H_{10}B_{1n}-J_{0}B_{2n}^*)
-(B_{1n}H_{10}-B_{2n}J_{0}^*)\right\}\right.\nonumber\\
&&\qquad\left.+{\hat S}\left\{(-J_{0}^*B_{1n}-H_{10}^TB_{2n}^*)
-(-B_{2n}^*H_{10}-B_{1n}J_{0}^*)\right\}\right.\nonumber\\
&&\qquad\left.+{\hat S}^{*}\left\{(H_{10}B_{2n}+J_{0}B_{1n})-
(B_{1n}J_{0}+B_{2n}H_{10}^T)\right\}\right.\nonumber\\
&&\qquad\left.-{\hat H}_2^T\left\{(-J_{0}^*B_{2n}+H_{10}^TB_{1n})
-(-B_{2n}^*J_{0}+B_{1n}H_{1}^T)\right\}|\epsilon^{[j]}\rangle
\delta_{ij}\delta_{mn}\right.\nonumber\\
&&+\left.\langle\psi^{[mi]}|{\hat H}_2\left\{(H_{10}B_{2j}
+J_{0}B_{2j}^*)-(B_{2j}H_{1}-B_{2j}J_{0}^*)\right\}\right.\nonumber\\
&&\qquad\left.+{\hat S}\left\{(-J_{0}^*B_{2j}+H_{10}B_{2j}^*)-
(B_{2j}^*H_{10}+B_{2j}^TJ_{0}^*)\right\}\right.\nonumber\\
&&\qquad\left.+{\hat S}^*\left\{(H_{10}B_{2j}-J_{0}B_{1j}^T)-
(B_{1j}J_{0}+B_{2j}H_{10})\right\}\right.\nonumber\\
&&\qquad\left.-{\hat H}_2^T\left\{(-J_{0}^*B_{2j}-H_{10}B_{1j}^T)-
(B_{2j}^*J_{0}-B_{1j}^TH_{10})\right\}|\epsilon^{[n]}\rangle
\delta_{ij}\delta_{mn}\right.\nonumber\\
&&\left.+h.c.\right],\label{supercharge of parity-like c2/z2}
\end{eqnarray}
where $m,n=1,2$ and $i,j=3,4$, and we define $B_{1m}$ as
$H_{1(2m)}+iH_{1(2m+1)}$, $B_{2m}$ as $J_{2m}+iJ_{2m+1}$, $B_{1i}$ as
$H_{1(2i)}+iH_{1(2i+1)}$ and $B_{2i}$ as $S_{2i}+iS_{2i+1}$. 

%%%%%%%%%%%%%%%%%%%
%\cite{Witten},\cite{Polchinski},\cite{BFSS},\cite{Hoppe},
%\cite{Horava-Witten},\cite{Candelas},
%\cite{Danielsson-Ferreti},\cite{Kachru-Silverstein},\cite{S.J.Rey},
%\cite{Z-orbifold},\cite{Candelas},\cite{Banks-Motl},
%\cite{Nilles},\cite{Katsuki-Kobayashi},\cite{Senda-Sugamoto},\cite{T^6},\cite{Kachra-Lawrence-Silverstein},\cite{Iso}.

%%%%%%%%%%%%%%%%%%%%%%%%%%%%%%%%%%%%%%%%%%%%%%%%%%%%%%%%%%%%%%%%%%%%%%%%%%%% 

%%%%%%%%%%%%%%%

\end{document}